\documentstyle[11pt,epsf,aaspp4]{article}

\def\etal   {{et~al.}\ }

\def\msun{{\rm\,M_\odot}}

\def\zdot{{\rm\,Z_\odot}}
\def\vol#1  {{{#1}{\rm,}\ }}
\def\lya{{\rm Ly}$\alpha$\ }

\def\etal{et al.\ }

\newcount\refno
\newcount\rfno
\def\eq{$^{\the\refno\ }$\advance\refno by 1}
\def\ad{\advance\rfno by 1}

\def\clock{\count0=\time \divide\count0 by 60
     \count1=\count0 \multiply\count1 by -60 \advance\count1 by \time
     \number\count0:\ifnum\count1<10{0\number\count1}\else\number\count1\fi}

\begin{document}
\title{Metal Enrichment and Temperature of the Intergalactic Medium}
\author{Renyue Cen\altaffilmark{1} and 
Greg L. Bryan\altaffilmark{2,3}}
\altaffiltext{1} {Princeton University Observatory, Princeton University, Princeton, NJ 08544; cen@astro.princeton.edu}
\altaffiltext{2} {Dept. of Physics, MIT; gbryan@mit.edu}
\altaffiltext{3} {Hubble Fellow}

\begin{abstract}

Hydrodynamic simulations of \lya clouds based on {\it ab inito}
cosmological models have produced results that are in broad agreement
with observations.  However, further analyses have revealed that, with
progressively higher numerical resolution, the median or cutoff line
width of the simulated \lya clouds (i.e. the Doppler parameter)
appears to converge to a value significantly below what is observed at
$z \sim 3$ (by about a factor of $1.5$).  These convergence
test simulations do not
include feedback from star formation. 
Given the observed metallicity in the Lyman alpha clouds
we suggest that supernovae, 
which presumably polluted the IGM with metals, may have
deposited a sufficient amount of energy in the IGM to reconcile the
theory with observations.  Simple arguments immediately narrow the
redshift range of pollution down to $4<z_{dep}<9$.  
It seems quite
certain that dwarf and sub-dwarf galaxies with total 
masses in the range
$10^{6.5-9.0}\msun$ have to be largely responsible for the pollution.
Furthermore, it is implied that either star formation is very
efficient or metal yield is very high for these early dwarf
galaxies, if the mean metallicity in the universe at $z=3$ is
as high as $<Z>=10^{-2}\zdot$.
Finally, assuming 
the specific supernova heating energy is proportional to
the metallicity of a gas, we note that the picture proposed here
would be consistent with supernovae being
the apparently needed heating source for intra-cluster gas 
if the required 
heating of the intra-cluster gas is no greater
than $1$keV per particle.

\end{abstract}

\keywords{Cosmology: large-scale structure of Universe
-- galaxies: formation
-- hydrodynamics
-- intergalactic medium
-- supernovae: general}

\section{Introduction}

Recent cosmological hydrodynamic simulations have shifted the paradigm
of \lya cloud formation.  Several independent groups have consistently
shown that \lya clouds are an integral part of the cosmic structure,
resulting from the gravitational growth of density fluctuations on
small-to-intermediate scales at $z\sim 2-4$ ($\sim 100$ kpc to a few
comoving megaparsecs) with most of the clouds in the form of
filamentary structures of moderate overdensity ($\delta_\rho \sim
1-10$) (Cen \etal 1994; Zhang \etal 1995; Hernquist \etal 1996;
Miralda-Escud\'e \etal 1996;
Wadsley \& Bond 1996; Cen \& Simcoe 1997;
Theuns \etal 1998;
Bryan \etal 1999; Dav\'e \etal 1999).

While the best agreement between model computations and observations
is seen in the column density distribution, a close examination
reveals that the Doppler width distribution shows a disturbingly large
discrepancy.  Specifically, Bryan \etal (1999) and Theuns \etal (1998)
have done convergence tests and shown that higher resolution
simulations predict a median Doppler width $b_{med,comp}\sim 20$km/sec
versus the observed $b_{med,obs}\sim 30$km/sec.  A comparable
difference is found between the low cutoff width of simulated and
observed clouds.  If this is interpreted as being due to a difference
in temperature, it means that the computed clouds are colder than the
real ones by approximately a factor of two, translating to a temperature
deficit of approximately $10^4~$K in the computed low column density clouds.

There are several ways to raise the temperature in the simulations to
bridge the gap.  First, increasing the baryonic density $\Omega_b$
allows more photoelectric heating and thus a higher IGM temperature,
with the scaling being $T\propto (\Omega_b h)^{1/1.7}$ (Hui \& Gnedin
1997), where $h$ is the Hubble constant in units of $100$ km/sec/Mpc.
Second, reionizing helium at a lower redshift will help somewhat (Hui
\& Gnedin 1997; Haehnelt \& Steinmetz 1998).  Third, Compton heating
by the high energy background may be important, if most of the
observed background at $z=0$ is produced at high redshift ($z>4$)
(Madau \& Efstathiou 1999).  
Oh (2000) has suggested that
X-rays from the first star clusters could provide this high energy
X-ray background.
Fourth, radiative transfer effects, which
are neglected in the simulations, may provide additional heating (Abel
\& Haehnelt 1999).
Finally,
photoelectric heating by dust grains (Nath, Sethi, \& Shchekinov 1999)
provides yet another mechanism and is more closely related to 
what is proposed in this paper. 

In this {\it Letter} we examine the possibility that feedback due to
early star formation in dwarf galaxies with masses $M\le 10^9\msun$
forming in the redshift range $7\le z_{f} \le 15$ may have played a
significant role in changing the thermodynamic state of the
intergalactic gas at moderate density.
The idea of supernova (SN) feedback into IGM 
can be traced back at least two and a half decades 
(Schwarz, Ostriker, \& Yahil 1975)
and many authors have subsequently addressed this
issue in various contexts 
(e.g., Ikeuchi \& Ostriker 1986;
Couchman \& Rees 1986;
Silk, Wyse, \& Shields 1987; 
Tegmark, Silk, \& Evrard 1993;
Shapiro, Giroux, \& Babul 1994).
While the basic physical process is by no means new, 
we address its effect in the new context
of the temperature of the \lya clouds.
In particular, we adopt the standard picture of re-ionization
due to radiation from quasars or massive stars, but argue that a
subsequent (or slightly overlapping) phase of energy ejection
from galaxies provides an additional source of heat for the IGM.

\section{Metal Enrichment and IGM Temperature}

Recent observations indicate that the metallicity in \lya forest
absorption lines with column density $N_{HI}\ge 3\times
10^{14}\;\hbox{cm}^{-2}$ is $\sim 10^{-2}\; \zdot$
at redshift $z\sim 3$ (Tytler \etal 1995; Songaila \& Cowie 1996).  We
show the possible implications of this observed metallicity on the
temperature of the \lya clouds.
In order to estimate the heating due to the supernovae which enriched
the IGM, we use the well measured carbon abundance to parameterize the
specific thermal energy increase:
\begin{equation}
e_{SN} = \frac{\eta E_{SN}}{M_{gas}}
       = \frac{\eta E_{SN}}{M_C} \frac{M_C}{M_{gas}}, 
\end{equation}
where $M_C$ is the mass of carbon ejected by one supernova; $M_{gas}$
is the total gas mass polluted by one supernova; $E_{SN}$ is the total
energy output of one supernova; $\eta$ is the fraction of that energy
that is eventually deposited in the IGM in the form of thermal energy.
It is useful to rewrite the term $M_C/M_{gas}$ as follows:
\begin{equation}
\frac{M_C}{M_{gas}} = 
\left(\frac{M_H}{M_{gas}}\right) 
\left(\frac{M_C}{M_{H}}\right) =
f_H \left[\frac{M_C}{M_H}\right]_\odot 
\left[\frac{C}{H}\right],
\end{equation}
where $f_H=0.76$ is the fractional mass in hydrogen;
$\left[{M_C/M_H}\right]_\odot=4.0\times 10^{-3}$ is the standard
ratio of carbon mass to hydrogen mass at solar abundance; $[C/H]$ is
the ratio of carbon number density to hydrogen number density of the
gas in solar units.  The temperature increase seen at redshift $z$
corresponding to this specific energy is given by:
\begin{eqnarray}
T_{SN}(z)&=&\frac{(\gamma - 1) \mu m_p e_{SN}}{k} 
        \left( \frac{1+z}{1+z_{dep}} \right)^2,
\end{eqnarray}
in the limit that the metal rich gas is deposited at redshift
$z_{dep}$ and adiabatic cooling dominates thereafter
(which may be a reasonable assumption since the IGM under consideration
has about the mean density).  Here $\mu$ is
the mean mass per particle in units of the proton mass $m_p$; $k$ is
the Boltzmann constant; $\gamma = 5/3$ is the ratio of specific heats.
Combining equations (1-3) we obtain the following expression
\begin{eqnarray}
T_{SN}(z)&=& 1.3 \times 10^4 
	\left( \frac{E_{SN}}{1.2\times 10^{51}{\rm erg}} \right)
	\left( \frac{M_C}{0.2\msun} \right)
	\left( \frac{\eta}{0.1} \right)
	\left( \frac{[C/H]}{3\times 10^{-3}} \right)
	\left( \frac{1+z}{1+z_{dep}} \right)^2 {\rm K}.
\end{eqnarray}
Here we have denominated $E_{SN}$ and $M_C$ by their respective,
reasonable values, following Woosley \& Weaver (1995); $M_C$ is
computed by averaging
metallicity yields over a Salpeter IMF.  We assume that most of the
carbon present comes from type II SNe, which, given the high
redshift of the enrichment, seems reasonable.

It is clear from equation (4) that, if the metallicity [C/H] lies in
the range $10^{-3}-10^{-2}\zdot$, feedback from SNe can
increase the temperature of the IGM by the required $\sim 10^4$ K
(Schaye \etal 1999; Bryan \& Machacek 2000; Ricotti, Gnedin \& Shull
2000; McDonald \etal 2000).

\section{Discussion}

When did the metal and energy deposition into the IGM from stellar
systems take place?  To answer this question in Figure 1 we plot the
required $\eta$ (left vertical axis) versus the deposition epoch
$z_{dep}$ (solid curve), such that the IGM temperature at $z=3$
(equation 4) is $2\times 10^{4}$K as observed in moderate density
regions (McDonald \etal 2000), where $E_{SN}=1.2\times 10^{51}$erg,
$M_C=0.2\msun$ and $[C/H]=3\times 10^{-3}$ are used.  Since $\eta$
cannot exceed unity, it is clear that energy and metals must have been
deposited at $z_{dep}<9$ in these regions.  The dashed curve shows the
corresponding required minimum initial temperature of the IGM (right
vertical axis) at $z_{dep}$, under the assumption that there is no
additional cooling or heating.  Other, associated cooling (primarily
hydrogen recombination cooling) will further reduce the temperature,
which would reduce $z_{dep}$.

What kind of stellar systems are likely to be responsible for the
contamination of metals and energy, if $z_{dep}$ is confined to the
range $3 < z_{dep} < 9$?  To answer this question, we adopt a very
simple, empirical model for the feedback.  We assume that galaxies of
a given mass form at $z_f$ and eject their metals and energy
instantaneously.  The exact mechanism for the ejection is likely to be
complicated and spherical blastwave models are inappropriate 
(e.g. Mac Low \& Ferrara 1999). 
In fact, it's not even clear if shock waves are
the primary mechanism for dispersal; other possibilities include
merging, turbulence, bulk flows and radiation pressure.  Instead, we
simply assume a constant dispersal velocity, $v_{disp}$ = 300 km/s.
In Figure 2 we plot the comoving separation of
halos as a function of halo mass, using Press-Schechter theory (Press
\& Schechter 1974).  Four curves are shown corresponding to
$z_{dep}=(3,5,7,9)$, with each curve being intersected by a horizontal
line to indicate the maximum distance a parcel of material could
travel with a dispersing speed $v_{disp}=300~$km/s within $1/2$ of the
respective Hubble time. 
We adopt a cosmological
constant dominated cold dark matter model with $H_0=65$km/sec/Mpc,
$\Omega_M=0.3$, $\Omega_\Lambda=0.7$ and $\sigma_8=0.90$ (Ostriker \&
Steinhardt 1995). 
Note that the corresponding halo formation
redshifts are somewhat higher due to the lag between the epoch of halo
formation and epoch of SN metal/energy deposition, with
$z_f=(5.4,8.5,11.7,14.9)$, respectively.  
We see that halos with masses $M_{halo} \le (30, 7,
1.8, 0.5) \times 10^8 \msun$ have to be responsible for injecting
metals and energy into the IGM   
at $z_f=(5.4,8.5,11.7,14.9)$,
respectively (it is assumed that the halo formation time and SN
injection time coincide while the energy deposition time in the
moderate density region is 1/2 Hubble time later).  Also shown in
Figure 2 are vertical arrows which indicate the minimum mass of halos
below which more than half of the gas cannot condense into the halos
due to the effect of gas pressure, at the four epochs (Gnedin 2000).
This further reduces the allowed range of halo masses which can heat
the IGM, constraining the epoch of formation/injection to be $z_{f}
\ge 7$.  It is fairly safe to conclude that dwarf and sub-dwarf
galaxies with masses in the range $10^{6.5-9.0}\msun$, forming in the
redshift range $z_f=7-15$ have taken a leading role in polluting the
IGM with metals and energy.


If the dwarf galaxies in the indicated redshift ranges are responsible
for the pollution, what is the implication for the star formation in
them?  Assuming the mean metallicity in the IGM is $<Z>$, we have
the following relation:
\begin{eqnarray}
Y f_* = {<Z>\over f_{col}},
\end{eqnarray}
where $Y$ is the overall yield of metals,
defined as the mass of metals
produced per mass of stars created,
$f_*$ is the star formation
efficiency of baryons within the collapsed halos (since SNe
from the first generation of stars in the dwarf galaxies may
completely disrupt the dwarf galaxies themselves, $f_*$ is essentially
the star formation efficiency of a primordial gas) and
$f_{col}$ is the fraction of mass in these dwarf galaxies.
Note that neither $Y$ nor $f_*$ can exceed unity.  
We integrate the mass function from the vertical arrow to the
intersection point
between each curve and its respective horizontal line in Figure (2)
to obtain $f_{col}$.
We summarize the various quantities in Table 1, for 
$<Z> = 10^{-2}-10^{-3}\zdot$.
Clearly we see that if $<Z> \sim 10^{-2}\zdot$ and the yield $Y$
is conventional ($\sim 0.02$), the efficiency of star formation in
these early dwarf galaxies must be quite high.
Conversely, a higher
yield $Y$, which would have profound implications for the initial
stellar mass function, can alleviate some burden on the star formation
efficiency.  But if $Z\sim 10^{-3}\zdot$, then no particularly high
star formation efficiency or unconventional yield is required. 
It would be most economical
if most of the polluting dwarf galaxies formed at $z_f\sim 8$.


If SN injection at these redshifts is important, it is not
obvious how it can occur and still retain the excellent agreement
between the observed and predicted column density distribution
previously mentioned.  As was noted, however, we do not
necessarily require that SN winds be the primary metal/energy mixing;
i.e., we do not require SN winds to have a traveling speed of 300
km/s.  It is likely that various processes are at work, including
SN winds, turbulent gas motions, bulk flows, and the
interactions of stellar systems (such as mergers and tidal
interactions).  It should be noted that the assumed dispersing speed
of $v_{disp}=300~$km/s is quite generous, since both galactic wind
speed and the bulk velocity of gas at high redshift are likely to be
less than that.
It seems that detailed simulations will be required to address this
point since it is unlikely that simple spherical winds can
reproduce the observations of metal distributions at high redshift
(Ferrara, Pettini \& Shchekinov 2000; Theuns, Mo \& Schaye 2000).

A word on the effect of feedback on subsequent galaxy formation is
relevant here.  Originally suggested by Efstathiou (1992) as a
mechanism to suppress galaxy formation in small galaxies, detailed
simulations indicate that photoionization heating alone fails to solve
the over-cooling problem (e.g. Navarro \& Steinmetz 1997).  It has
been noted recently by Gnedin (2000) that the effect of pre-heated gas
(by photoionization in his case) on the accretion of gas onto galaxies
is larger (by a factor of $\sim 10$ in galaxy mass) than expected
based on a simple Jeans' analysis.  The feedback heating discussed
here increases the temperature of photonized gas by a factor of two
and so would raise the filtering mass to $\sim 1.5\times 10^{10}\msun$
at $z\sim 4$ (see Figure 3a of Gnedin 2000), a factor of $\sim 4$
higher than found in Quinn, Katz, \& Efstathiou (1996).
This may be sufficient to solve the over-cooling problem.
Another possible benefit is a
delay in the accretion of gas onto larger galaxies,
which may alleviate the angular momentum problem in spiral disks
found in simulations (e.g. Navarro \& Steinmetz 1997)

Comparison with previous work is useful.  Gnedin \& Ostriker
(1997), Gnedin (1998) and Cen \& Ostriker (1999) have studied the
issue of metal enrichment using direct numerical simulation including
heuristic galaxy formation and metal enrichment prescriptions, as does
the recent work by Aguirre \etal (2000) who focus on larger galaxies
at $z\sim 3$.  Our work is complementary in terms of both methodology
and scale/epoch 
in that we focus on the possibility of metal and energy
enrichment by smaller galaxies at higher redshift.  The work of Cen \&
Ostriker (1999) hints that the high redshift universe may not have
been sufficiently enriched in metals due, possibly, to the
lack of numerical resolution to adequately simulate the formation of
small galaxies at high redshift.  The results in Aguirre \etal (2000)
again indicate that large galaxies at relatively low redshift
may have difficulty in fulfilling the task of enriching
the universe uniformly, consistent with our results.  Our work is
similar to that of Gnedin \& Ostriker (1997) and Gnedin (1998) in that
both focus on smaller galaxies at high redshift, but using different
methods.  While our treatment cannot study the detailed metal/energy
enrichment process (i.e. we do not know how the metals get
out of galaxies), current simulations without a model for the
multi-phase interstellar medium have to make similar assumptions and
parameterizations (e.g., the yield parameter and wind speed).  What is
possible with current simulations is to explore the various possible
metal dispersing mechanisms including SN winds (Cen \& Ostriker
1999; Aguirre \etal 2000) and mergers (Gnedin \& Ostriker 1997; Gnedin
1998 among others), once metals do get out of galaxies.  In any case,
our treatment does not advocate to/require any specific metal/energy
dispersing mechanism, rather we only need to assume an average
dispersing velocity. 

The galaxy masses which we identify as responsible for the heating are
generally smaller than typical present-day galactic systems.  In fact,
the smallest systems which can cool and so form stars are around
$10^{6} M_{\odot}$ (Tegmark \etal 1997; Haiman, Abel \& Rees 2000).
These first stellar systems would correspond to the lower end of the
range of systems we discuss.  If, as recent simulations indicate
(Abel, Bryan \& Norman 2000; Bromm, Coppi \& Larson 1999), such
systems form predominantly massive stars, $\sim 100 M_{\odot}$, and
the energy and metal feedback expected from the resulting SNe
would indeed pollute much of the IGM.  On the other hand, some of
these same simulations show that the efficiency of such star formation
may be very small ($< 1\%$), in which case more massive (but still
dwarf) galaxies could play the role suggested in this paper.  There
are indications that their star formation efficiency may be
considerably larger (Machacek, Bryan \& Abel 2000).

\section{Conclusions}

We suggest that energy feedback from supernovae in dwarf or sub-dwarf
galaxies with masses in the range $10^{6.5-9.0}\msun$ forming at
$z_{f}=7-15$ could raise the temperature of the \lya clouds at $z\sim 3$
by $\sim 10^4$K, given the observed metals in the \lya clouds at $z\sim
3$.  This provides an alternative mechanism to bring the results of
recent hydrodynamic simulations into agreement with the observed 
Doppler width distribution of \lya clouds.

This scenario may be tested in a number of ways.  First, since metals
are a sign of heating, we would expect a correlation between
metallicity and temperature (i.e. Doppler width).  A search for this
effect should be restricted to low column density lines ($N_{HI} <
10^{14}$ cm$^{-2}$ at $z \sim 3$) where hydrogen line cooling is
unimportant.  Since it is difficult to detect metal lines in such
systems, a statistical search may be necessary (e.g. Cowie \& Songalia
1998).  Another test would be to examine the evolution of the
temperature of Ly$\alpha$ clouds at high redshift $z \ge 4$.  In
particular, the mean temperature should not increase with redshift as
rapidly as in the case with only photoionization heating and adiabatic
cooling.  This is because heating (and subsequent cooling) due to
contamination is likely to be continuous over the entire redshift
range.  A third prediction would be a larger spread in the temperature
of the gas at a given density, since heating would no longer depend
just on density (as it does for radiative heating) but instead would
occur suddenly.  This might show up as a larger scatter in the joint
$N_{HI}-b$ distribution for the Ly$\alpha$ forest.  Since we require
strong outflows from dwarf galaxies, a fourth suggestion would be to
attempt to directly observe these outflows.  This would be difficult
at the relatively high redshifts we predict, but may be possible if
the outflows are driven by strong starbursts.  
Finally, at low
redshift we predict that the metallicity in the low column density
Ly$\alpha$ forest clouds should remain approximately constant at $Z
\sim 10^{-2} - 10^{-3} Z_{\odot}$, since later, larger galaxies are
unlikely to be able to signficantly contaminate the regions further.
An interesting aside is that collisional ionization at earlier times
could be significant if energy feedback from SNe raised the IGM
temperature, and may significantly alter the reionization picture of
the universe.  This issue will be examined in detail subsequently.

We note that the SN heating scenario for the high redshift
IGM may be related to the apparently wanted additional 
heating of intra-cluster gas,
{\it if the latter is also due to SNe}.
If we take the apparently required specific energy per baryonic particle 
of $0.3$keV for the intra-cluster gas
(e.g., Lloyd-Davies, Ponman, \& Cannon 2000)
and simply assume that specific heating energy is linearly 
proportional to metallicity, we obtain
the specific energy per baryonic particle 
for the high redshift IGM to be 
$Z_{IGM}/Z_{ICM}\times 0.3$keV$=3\times 10^{-3}/0.3\times 0.3$keV$=3$eV,
not inconsistent with the results presented in this paper.
However, a specific heating 
energy per baryonic particle for the intra-cluster gas
greater than $1$keV is unattainable within our picture.

\acknowledgments
We would like to thank an anonymous referee for useful comments.
This research is supported in part by grants AST93-18185
and ASC97-40300.
Support for this work was also provided by NASA through Hubble 
Fellowship grant HF-01104.01-98A from the Space Telescope Science
Institute, which is operated by the Association of Universities for
Research in Astronomy, Inc., under NASA contract NAS6-26555.


\begin{figure}
\epsfxsize=5in
\centerline{\epsfbox{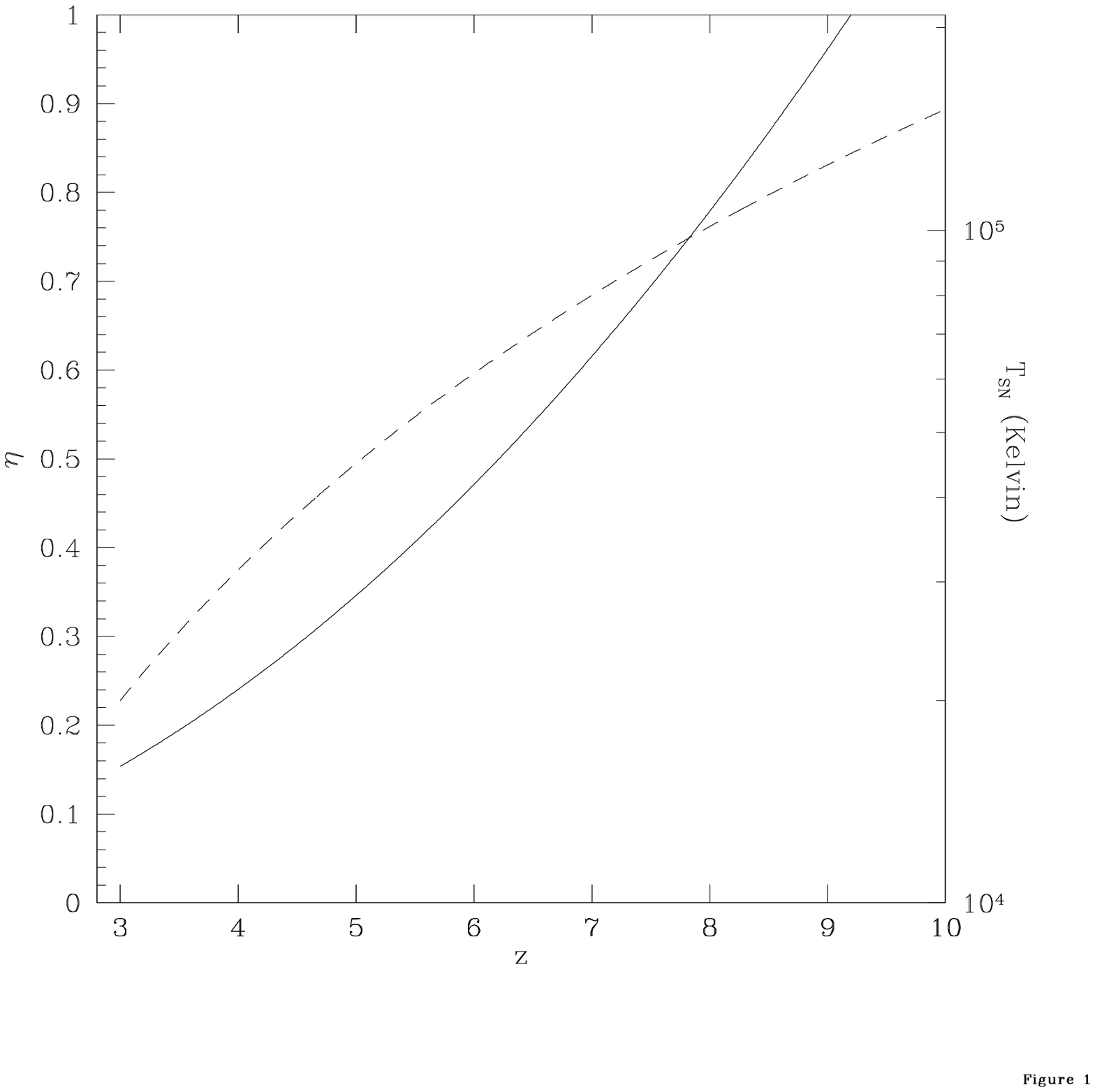}}
\caption{
Shows the required $\eta$ (left vertical axis) versus the
deposition epoch $z_{dep}$ (solid curve), such that the IGM
temperature at $z=3$ (equation 4) is $2\times 10^{4}$K for moderate
density regions (McDonald \etal 2000), where $E_{SN}=1.2\times
10^{51}$erg, $M_C=0.2\msun$ and $[C/H]=3\times 10^{-3}$ are used.  The
dashed curve shows the required minimum initial temperature of the IGM
(right vertical axis) at $z_{dep}$, under the assumption that there is
no additional cooling.  }
\label{fig1}
\end{figure}

\begin{figure}
\epsfxsize=5in
\centerline{\epsfbox{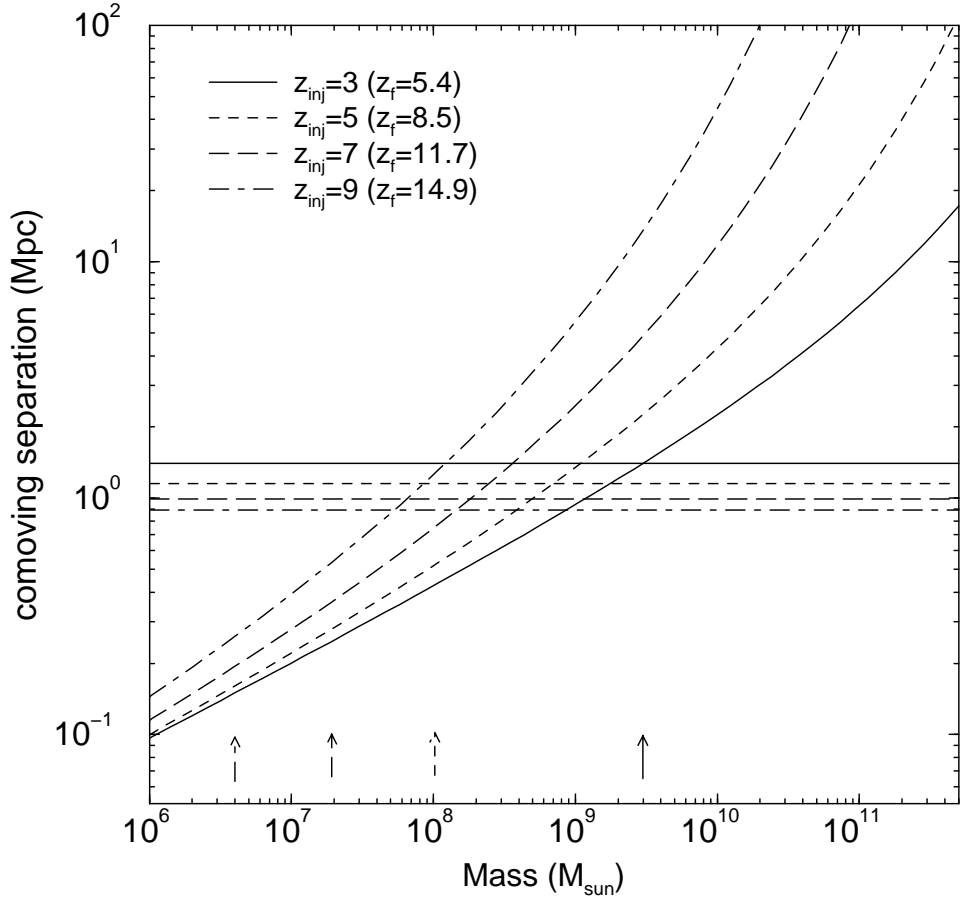}}
\caption{ 
The curves show, for four redshifts, the mean comoving
separation between collapsed objects as a function of their virial
mass for a $\Lambda$CDM cosmology.  Horizontal lines show the maximum
distance that a parcel of material with a velocity 300 km/s could
travel in 1/2 of the Hubble time at that redshift.  The vertical
arrows indicate the minimum mass of halos below which more than half
of the gas cannot condense into the halos, at the four epochs (Gnedin
2000).  }
\label{fig2}
\end{figure}

\begin{deluxetable}{cccccccccc} 
\tablewidth{0pt}
\tablenum{1}
\tablecolumns{12}
\tablecaption{Characteristics of Polluting Dwarf Galaxies} 
\tablehead{
\colhead{$z_{dep}$} &
\colhead{$z_{f}$} &
\colhead{Halo Mass} &
\colhead{Total mass fraction} &
\colhead{$Y f_*$\footnote{Assuming the mean metallicity $<Z>=10^{-3}-10^{-2}\zdot$}} &
\colhead{$f_*(Y=0.02)$}}

\startdata
$3$ & $5.4$ & $3.0\times 10^9-3.0\times 10^9$ & $0.0$ & $>1$ & $>1$ \nl 
$5$ & $8.5$ & $1.0\times 10^8-7.0\times 10^8$ & $0.030$ & $0.0067-0.067$ & $0.033-0.33$ \nl 
$7$ & $11.7$ & $2.0\times 10^7-2.0\times 10^8$ & $0.013$ & $0.015-0.15$ & $0.077-0.77$ \nl 
$9$ & $14.9$ & $4.0\times 10^6-6.0\times 10^7$ & $0.010$ & $0.020-0.20$ & $0.1-1.0$ \nl 
\enddata
\end{deluxetable}
\end{document}